\documentclass{aa}

\def\beq{\begin{equation}}

\def\arcsec{\hbox{$''$}}

\def\eeq{\end{equation}}

\def\percc{\hbox{cm$^{-3}$}}

\def\percmsq{\hbox{cm$^{-2}$}}

\def\H2*{\hbox{H$_2^*$}}

\def\arcmin{\hbox{'}}

\def\HI {\hbox{H~{sc i}}}

\def\H#1#2{\hbox{H~#1#2$\alpha$}}

\def\kms{~km~s$^{-1}$~}

\def\kmsns{~km~s$^{-1}$}

\def\eg{e.g.,~}

\def\etal{{et~al.~}}

\def\AMM{\hbox{NH$_3$}}

\def\NH3{\hbox{NH$_3$}}

\def\INH3{\hbox{$^{15}$NH$_3$}}

\def\NINH3{\hbox{$^{14}$NH$_3$}}

\def\vlsr{\hbox{$v_{LSR}$}}

\def\ICO{\hbox{$^{13}$CO }}

\def\HI{\hbox{H{\sc i }}}

\def\MOLH{\hbox{H$_2$}}

\chardef\isp="10 

\usepackage{graphicx}

\begin{document}

\title{An Alternate Estimate of the Mass of Dust in Cassiopeia A}

\author{T.~L.~Wilson\inst{1,2}
    \and W.~Batrla\inst{2}\thanks{Current address: Fachhochschule Hof, Abteilung M\"unchberg,
        Kulmbacher Strasse 76, D-95213 M\"unchberg, Germany}}

\institute{European Southern Observatory, K-Schwarzschild-Strasse
        2, 85748 Garching, Germany
    \and Max-Planck-Institut f\"ur Radioastronomie, Auf dem H\"ugel 69, 53121
        Bonn, Germany}

\offprints{twilson@eso.org }

\date{Received:FILL IN; accepted:FILL IN}

\authorrunning{T.~L.~Wilson \& W.~Batrla}

\titlerunning{Dust in Cassiopeia A}
\abstract{Recent observations of sub-millimeter continuum emission toward supernova remnants (SNR) have raised the question of whether such emission is caused by dust within the SNR itself or along the line-of-sight. Here we make a comparison of the image of sub-mm emission from dust with the integrated $J=1-0$ line emission from interstellar \ICO\ toward the SNR Cassiopeia A based on existing data. The cm and mm synchrotron emission from Cas~A has a rather symmetric, ring-like structure whereas both the sub-mm continuum and interstellar \ICO\ line emission are located mostly toward the south of the SNR. There is positional agreement for 3 of 6 maxima found in \ICO\ line and sub-mm continuum emission, with the weakest feature near the center of Cas~A and the other two features near the southeast and west edges of the SNR. For these three maxima, a comparison of masses determined from dust and \ICO\ data shows good agreement if we use the 450 $\mu$m dust absorption coefficient typical for diffuse clouds. There is also good agreement between the sub-mm dust temperature and the gas kinetic temperature from CO and \AMM. For the remaining sub-mm continuum peaks, one is outside the forward shock of the SNR. For the other two, one was not mapped in $^{13}$CO; for the other there is no $^{13}$CO emission. \HI\ absorption covers all of Cas~A, but the \HI\ column density may be too small to account for the sub-mm dust emission. Thus it is possible that one, or perhaps two of these sub-mm continuum peaks are located inside the SNR.  From lower resolution maps in CO lines, the SE and W features are the edges of extended clouds. Toward the cloud centers, the CO emission is more intense, but there appears to be less sub-mm dust emission. The differences between CO and sub-mm images may be caused a combination of the techniques used to produce the sub-mm maps and changes in cloud properties from center to edge. 
         }
\maketitle
\section{Introduction}

Large amounts of dust have been observed in a number of galaxies
with large redshifts (see e.g. Smail \etal \cite{sma},
Eales \etal \cite{eal}). Dust is produced in red giant stars and
ejected in winds, but Morgan and Edmunds (\cite{mo2}) argue that
this process is too slow to explain the presence of large amounts
of dust in the early universe unless the star formation rates are
very large. Morgan \etal(\cite{mo1}) have presented an image of
dust emission of Kepler's supernova remnant (SNR) and estimate
that 1~M$_\odot$ of dust is produced.  Dunne \etal(\cite{dun})
have estimated that Cassiopeia A has produced 2 to 4~M$_\odot$ of
dust. On the basis of this data and the data of Morgan
\etal(\cite{mo1}), Dunne \etal(\cite{dun}) have argued that large
amounts of dust can be produced in SNe. In the Cas~A image, much
of the dust emission is concentrated toward the southern part of
the SNR. Dunne \etal(\cite{dun}) had assumed that all of the dust which gives rise to sub-mm emission   
is contained in the SNR itself and was produced by the SNR itself. From a comparison of 24 $\mu$m, 70 $\mu$m and sub-mm dust images, Hines \etal (2004) noted that if the dust emitting at sub-mm wavelengths resides within the SNR, it must have very different properties from the dust emitting in the mid-IR.  
Batrla \etal(\cite{bat}), Troland \etal(1985), Wilson \etal(1993), Gaume \etal(1994) and  Liszt \& Lucas (1999) have
measured molecular absorption and emission toward the southern
part of Cas~A. These results give a gas kinetic temperature which is the same as the temperature reported by Dunne \etal(\cite{dun}). In fact all of these temperatures are about the same as the equilibrium temperature of cirrus clouds deduced from COBE satellite data (Lagache \etal 1998).  Thus, at least some of the sub-mm dust emission assigned to the SNR
by Dunne \etal(\cite{dun}) may arise in the
interstellar medium (ISM).

In this paper, we compare the spatial distribution of sub-mm dust emission (15\arcsec\ resolution) with \ICO\ emission (21\arcsec\ resolution), CO emission (28\arcsec\ resolution), with \HI\ absorption (7\arcsec\ resolution) toward Cas~A, and with CO maps (2.5\arcmin\ resolution) around Cas~A  to separate sub-mm dust emission from  interstellar clouds toward the SNR Cas~A from sub-mm dust emission arising from within this SNR.

\section{Comparison of Results}
\subsection{Qualitative Comparison}

\begin{figure*}[ht]
\label{850_CO}
{\includegraphics[scale=0.70]{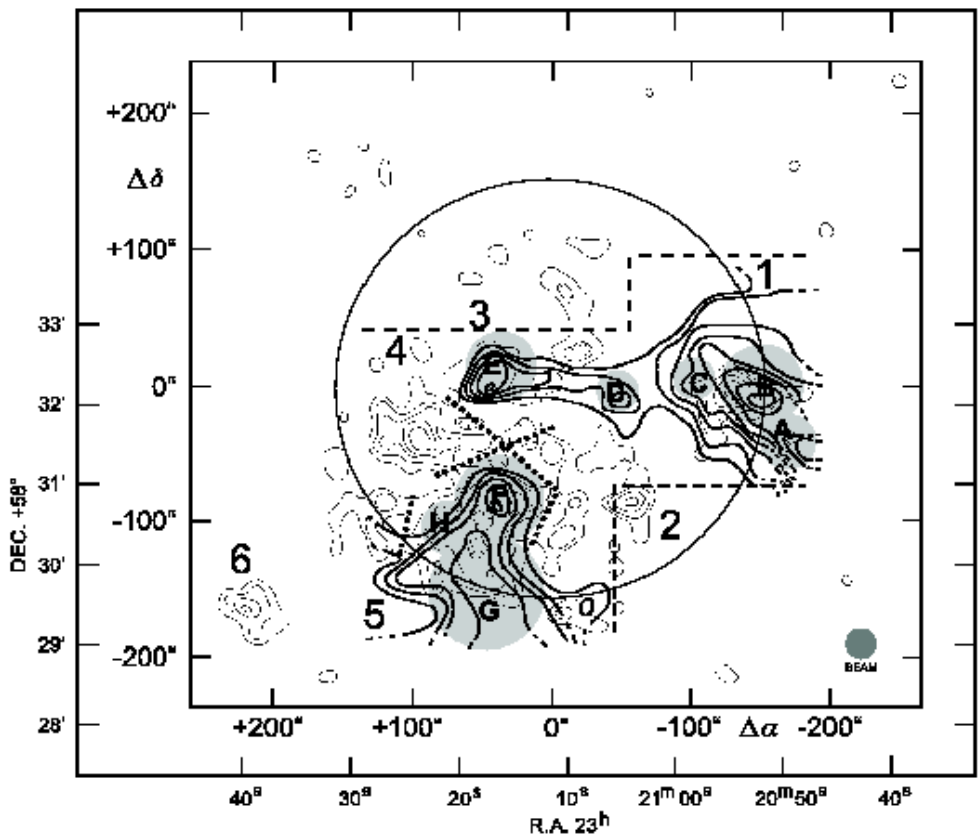}} \caption{Overlay of
the $J=1-0$ \ICO\ integrated line intensity, shown as thick
contours, starting at 4 K km s$^{-1}$, in steps of 2 K km s$^{-1}$ to 18 K km s$^{-1}$ (from Wilson \etal(\cite{wil})) on 450~$\mu$m sub-mm dust emission image, shown as thin contours (from Dunne \etal(\cite{dun})). The B1950.0 coordinates are taken from Wilson \etal (1993). See Fig.~2 for a map of a larger region in the CO $J=2-1$ line. 
 We have numbered the sub-mm dust emission maxima 1 to 6. The
\ICO\ clouds are labeled with capital letters A to H. Shaded circles
show the FWHP and location of \ICO\ clouds (from Wilson
\etal(\cite{wil})). The sub-mm dust peaks under E and F do not have higher contours for clarity. The thick dotted lines mark the boundaries where we have divided the sub-mm dust emission between peaks 2, 3, 4, and 5. We used these as the boundaries of integrations to determine the masses of individual regions.}
\end{figure*}

The \ICO\ J=1--0 line data for our comparison were taken with the
IRAM 30-m radio telescope on Pico Veleta. The angular resolution
was 21\arcsec, the observing method position switching with the
reference position 60\arcmin\ north of the center of Cas~A. The
most important point is that all of the \ICO\ emission, even from
very extended clouds, was recorded. Details are to be found in
Wilson \etal(\cite{wil}).

The sub-mm data presented by Dunne \etal(\cite{dun}) were taken with a
15\arcsec\ angular resolution. Observational details are given by Loinard \etal (2003). To suppress atmospheric emission,
one should difference the signal between two regions of the sky separated by a number of beamwidths; images made from this data
were restored using the procedures described by Johnstone 
\etal(\cite{john}).  Johnstone \etal(\cite{john}) present a number of data reduction techniques; that used by Dunne \etal(\cite{dun}) is from Emerson \etal (1979 hereafter EKH). All restorations tend to suppress the signal from structures with sizes larger than twice the maximum chopper throw, and also increases the noise in the corresponding Fourier components. According to Johnstone \etal (2000), the EKH method produces lower quality images near map edges. Loinard \etal(2003) did not specify their chopper throw, but the maximum throw used at JCMT is 120\arcsec.  The restoration process produces sub-mm emission maps at 850 and 450 $\mu$m. A map of the synchrotron emission (see \eg Wright \etal 1999), scaled in frequency, was subtracted from these images. The final 450 $\mu$m image is given in Fig.~4 of Dunne \etal(\cite{dun}). Although the error beam of the telescope is larger at 450 $\mu$m and atmospheric quality is worse than at 850$\mu$m, the sub-mm dust intensity is larger, so the sub-mm dust emission is less affected by possible residual synchrotron emission from the SNR. All our comparisons are made with the 450 $\mu$m map.

In Fig.~1 we show an overlay of the J=1--0 \ICO\ integrated line
intensity image (Fig.~3 of Wilson \etal \cite{wil}), shown as
thick contours with the 450 $\mu$m dust image (Fig.~4 of Dunne
\etal \cite{dun}) shown as thin contours. The \ICO\ clouds are
labelled using capital letters taken from Wilson
\etal(\cite{wil}). Shaded circles show the gaussian Full Width to Half
Power (FWHP) sizes and locations of \ICO\ clouds, as cataloged by Wilson
\etal(\cite{wil}). 
The thin circle of radius 153\arcsec, centered on offset ($\Delta \alpha, \Delta \delta$=0\arcsec, 0\arcsec) marks the
location of the {\it forward shock} from the SNR.  The dust is thought to form inside this shock, but sub-mm dust peak 6 (at (216\arcsec, -163\arcsec)) clearly lies outside.  The shaded circle at
the right bottom is the half power beam width for the \ICO\
data, 21\arcsec. The box containing the offset coordinates is the full size of Fig.~4 of Dunne \etal(\cite{dun}). The dashed lines show the limits of the region in which the \ICO\ $J=1-0$ line was fully sampled. 

\begin{figure}[h]
\label{COlisluc}
{\includegraphics[scale=0.4]{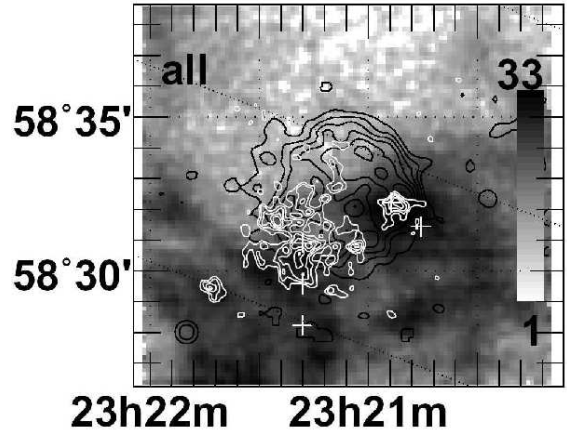}}\caption{ Adapted from Liszt \& Lucas 1999; the coordinates are B1950.0.  The gray scale is the $J=2-1$ CO emission toward Cas A, summed from V=-30 to -50 km s$^{-1}$. The wedge on the right side of the image, with numbers from 1 to 33, is the scale for integrated intensity in units of K$\cdot$km s$^{-1}$. The white crosses indicate where spectra were taken with longer integrations by Liszt \& Lucas (1999). The range of radial velocity is basically the same as that used to form the \ICO\ image, shown as thick contours in Fig. 1. The angular resolution of the CO data is 28\arcsec. The black contours are the 140 GHz continuum from the SNR Cas A measured by Liszt \& Lucas 1999, with angular resolution 44\arcsec. This is dominated by synchrotron emission. We have superposed the sub-mm dust emission contours from Fig.~4 of Dunne \etal(\cite{dun}) as white contours. This image shows that most of the CO is located south of Cas A, as is the sub-mm dust emission. The \ICO\ extent (Fig.~1) is smaller than that of the CO because of lower optical depth. } 
\end{figure}

To show the full extent of CO emission toward Cas A, in Fig. 2, we show the gray scale image of the $J=2-1$ line of CO superposed on contours of the continuum emission from Cas A at 140 GHz (Liszt \& Lucas 1999). This 28\arcsec\ resolution image shows more intense CO emission to the SE and SW of the center of Cas A, but little to the north. 
The optical depth of the CO lines is $>$5 (Troland \etal (1985), but the isotopic ratio of CO/\ICO\ for these clouds is 1/60 (Wilson \& Rood 1994), so the \ICO\ lines are optically thin, and so the \ICO\ line emission traces H$_2$ column density. In addition, the intensity of the \ICO\ line emission is always small and the \ICO\ emission always has a smaller extent than CO emission (see \eg Section 14.8.1 in Rohlfs \& Wilson 2003). Thus even though the \ICO\ data in Fig 1 (between the dashed lines) does not cover the northern part of Cas A, one can be certain that there is no \ICO\ emission from this region. In Fig. 3, we show the lower resolution CO $J=1-0$ line maps of Troland \etal (1985). There is emission over a wide region to the south of Cas A. We have added  a thick square box to the figure of Troland \etal (1985) to show the maximum area mapped by Dunne \etal (2003) at 850 $\mu$m. The region in Fig.~1 is 20\% smaller. From the \ICO\ and CO data, it is clear that the CO emission measured toward Cas A arises from the edges of larger clouds. These clouds have small peak temperatures, so are low density interstellar clouds that contain no embedded high mass stars. From these data it is evident that:

\noindent
(1) there is sub-mm dust emission to the south of Cas A, but {\it not} toward the northern part of Cas A. 

\noindent
(2) The synchrotron radio continuum emission from Cas A is highly symmetric (Fig.~2 and Wright \etal 1999). 

\noindent 
(3) Most of the molecular clouds are found toward the south of Cas A. 

\noindent 
(4) There is a least one sub-mm dust emission peak, clump 6, outside the forward shock of the SNR. This cannot be a projection effect, but the location of clump 3 might be.

There is spatial overlap of three sub-mm dust maxima of Dunne \etal (2003) with groups of \ICO\ peaks toward the SE of the SNR (clouds F, G, and H), toward the center (cloud E), and toward the west(clouds A, B and C). There is very little CO or sub-mm dust emission toward the north of Cas A. This leads one to suspect a possible connection between the CO clouds in the ISM and the sub-mm dust emission measured by Dunne \etal (2003). From Fig.~1 the sub-mm dust emission decreases outside the forward shock. Assuming that the sensitivity and response of the sub-mm dust measurements of the region shown as a box in Fig. 3 are uniform, and if there is a proportionality between CO, \ICO\ and sub-mm dust emission in interstellar clouds, we must explain the decrease in sub-mm dust emission to the south of Cas A. There are two possible causes:

(1) A change in cloud properties from the edge to cloud center. As is widely accepted, molecular clouds consist of many small clumps (see \eg Tauber 1995 and references therein). Toward the edges one expects fewer clumps, resulting in prominent fine scale structure. Toward the cloud centers, however, there are many overlapping small clumps, which would produce an apparently smooth cloud structure. The beam-chopped measurements of sub-mm dust emission would suppress the apparently smoother structure in the cloud centers (see \eg Johnstone \etal 2000). Such an effect would reduce the intensity in the southern part of the map of Dunne \etal (2003). This effect is enhanced since the sub-mm images were reconstructed using the EKH technique. 

(2) Less likely is an interaction between the SNR and CO clouds. From Spitzer 24 $\mu$m data, Hines \etal(2004) noted evidence for such an interaction with a wispy filament of CO (Liszt \& Lucas 1999 data) to the north of Cas A. Hines \etal (2004) report no evidence for an interaction with CO clouds to the south however.   

\begin{figure*}[h]
\label{COtroland}
{\includegraphics[scale=1.0]{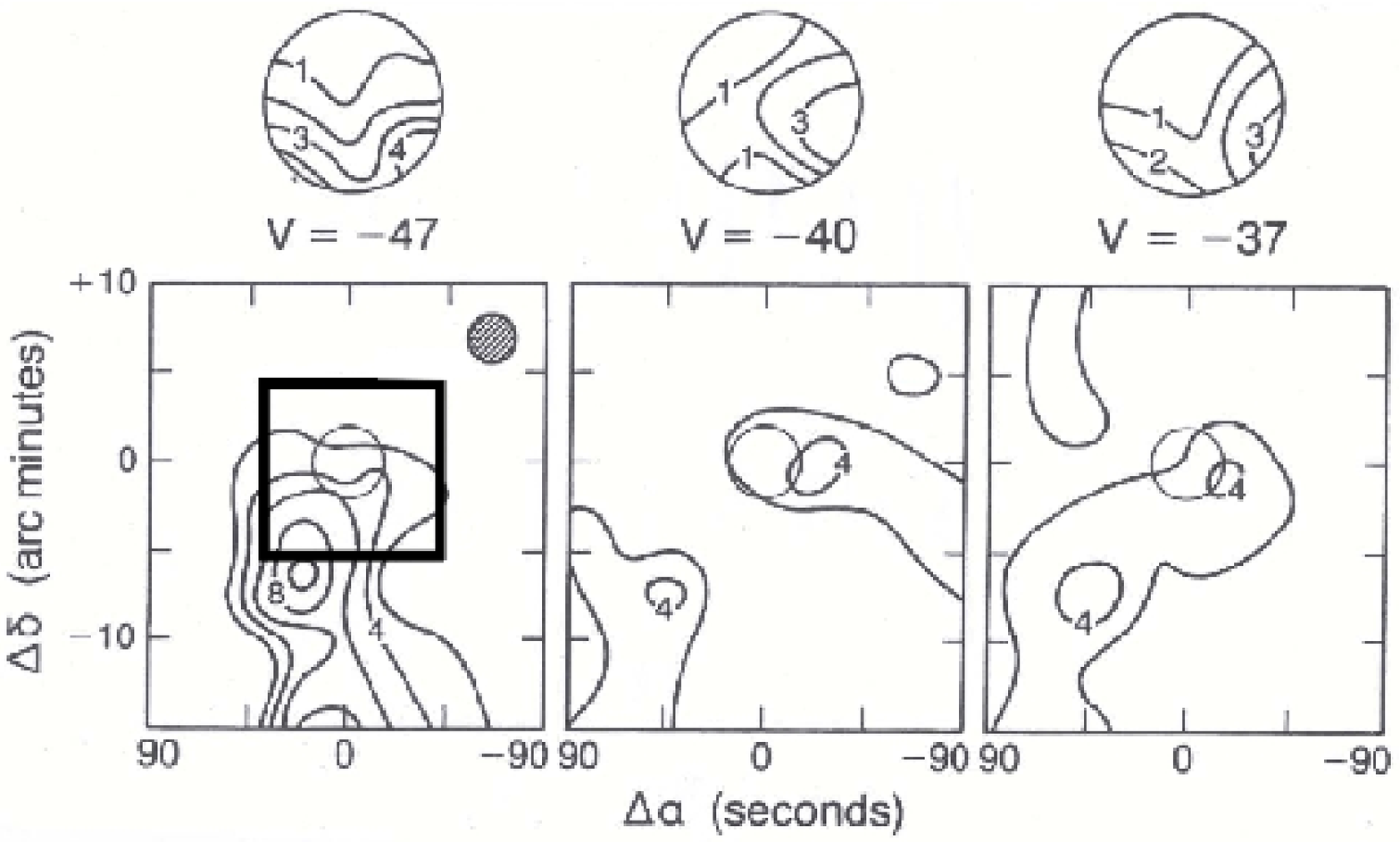}} \caption{ Adapted from Troland \etal 1985 (resolution 2.5\arcmin). The three panels show the radial velocities which contribute to the CO emission. In the upper three sketches, the coverage of the CO emission over the face of Cas A is shown. As in Fig 2, most of the CO emission is to the south of the center of Cas A. In each lower panel the thin circle at offset (0,0) marks the edge of the continuum emission from Cas A. The FWHP beam is shown in the upper right of the V=-47 km s$^{-1}$ panel. The contour units are peak line intensity in K. The thick square box in this panel shows 1.2 times the boundaries of the 450 $\mu$m sub-mm dust continuum map of Dunne \etal 2003. These data show that the clouds covering the southern part of Cas A are the edges of larger molecular clouds.}
\end{figure*}

\subsection{Quantitative  Masses and Column Densities}
\subsubsection{\HI Clouds} 

There is no \ICO\ cloud coincident with sub-mm dust clump
\lq4\rq. Clump \lq2\rq\ was included in the \ICO\ mapping.  There is a significant 
amount of \HI  absorption
toward  most of   Cas A. Bieging \etal (1991) have imaged the \HI with an angular resolution of 7\arcsec. The cloud they have labeled as $'A'$ (maximum optical depth, $\tau$
of 3 at \vlsr=-49.7\kmsns) overlaps with the sub-mm dust peak we labeled as  \lq2\rq. The \HI cloud that Bieging \etal (1991) labeled $'D'$ ($\tau$=2 and \vlsr=-43.9 \kmsns) overlaps with the sub-mm dust peak we labeled as \lq1\rq. In addition, Bieging \etal (1991) find a cloud labeled $'B'$ that covers the entire continuum source. 

From, \eg Rohlfs \& Wilson (2003)
the relation between \HI  line parameters and \HI\ column density, N, is  ${\rm N} =1.822 \times 10^{18} {\rm T_s} \, \int \tau {\rm d } v$. Here $\tau$ is the \HI  optical depth, $v$ is the radial velocity in \kms and T$_{\rm s}$ is the spin temperature. From the data of Bieging \etal (1991), it is difficult to determine the value of T$_{\rm s}$. If the \HI\ spin temperatures are $\sim$20 K, and the linewidths are 3.3 \kms (the average $\Delta$V of the \ICO), the column densities of the \HI  clouds are $\sim$10$^{20}$\percc. This value is $<$10\% of the average column density estimated for the sub-mm dust peaks. Our estimate of the \HI  column density is a minimum. It is possible that the values of T$_{\rm s}$ and $\Delta$V are larger, so that the column density of the \HI  would be larger. It is unclear whether the column density of \HI could be as much as a factor of 10 larger. If it were, one could assign {\it all} of the sub-mm dust emission to the interstellar medium. For this, one would need a secure determination of the spin temperature of the \HI. Until this is done, one must conclude that while some of the sub-mm dust peaks measured by Dunne \etal (2003) {\it are} interstellar, \lq2\rq\ and
\lq4\rq\ could be located 
inside the SNR.

 \subsubsection{\ICO\ Clouds}

From the \ICO\ data
the beam averaged column densities of these groups of clouds are 6.6 10$^{21}$ \percmsq\ (A, B and C), and 2.4 10$^{21}$ \percmsq\ (F, G and H).  
As noted by Wilson \etal(\cite{wil}) and Batrla
\etal(\cite{bat}), the interstellar clouds toward Cas~A show
indications of clumping on scales down to the respective beam
sizes. Thus the beam-chopped measurements of sub-mm dust emission would have recorded these clouds.

In Table 1, we have taken the measured cloud FWHP sizes and average H$_2$ densities from Wilson \etal(\cite{wil}); these are shown in Fig.~1 as shaded circles. Wilson \etal (1993) had used the average H$_2$ densities (Col.~(4) of Table 1), with the FWHP gaussian sizes (Col.~(2) of Table 1) to determine masses. Walmsley \& Panagia (1978) favor a procedure that uses radii obtained from the assumption of a spherical source shape. In this model, if the beam size is much smaller than the source size, the diameter is 1.47 times the FWHP size (see \eg the Appendix of Mezger \& Henderson (1967)). In addition, we used a factor of 1.36 to account for the mass in helium.  The gas masses are given in Col.~(5) of Table 1; from a gas-to-dust mass ratio of 100 we obtain dust masses in Col.~(6) of Table 1 and  Table 2. 

\subsubsection{Dust Clouds}

To determine the dust masses of individual clumps from the integrated sub-mm intensities, we used the 450 $\mu$m contours in Fig.~4 of Dunne \etal (2003). We measured the integrated intensities by digitizing contours, converting to an {\it autocad} file to measure the areas enclosed by each of the contours. This does not include contours below 3$\sigma$, since these were not in Fig.~4 of Dunne \etal (2003). Although an underestimate of the actual total integrated flux, this is adequate to obtain the relative flux densities for the clumps. We divided the contours along the depressions to separate clumps (dotted lines in Fig.~1) to obtain the 450 $\mu$m integrated flux densities for the individual clumps. From these we obtained fractional values; when multiplied by the total masses (tabulated in Table 1 of Dunne \etal 2003) we obtained the dust masses for the clumps. 

\subsubsection{Comparison of \ICO\ and sub-mm dust masses}

We compare the dust masses from the
450~$\mu$m and \ICO\ data for the overlapping clouds in Table 2.  
To compare the masses from sub-mm dust emission with \ICO\ line emission, we must have 450~$\mu$m absorption coefficients for the conversion from thermal dust intensity to mass, that is, values of $\kappa_{450}$. Dunne \etal (2003) argued that they must use a large value, $\kappa_{450}$=1.5 m$^2$ kg$^{-1}$, which corresponds to an {\it amorphous} or clumpy dust structure, to avoid having too much dust produced in Cas A. We used this value to obtain the masses in Col.~3 of Table 2. The value for {\it newly formed} dust, in \eg reflection nebulae or planetary nebulae, 0.88 m$^2$ kg$^{-1}$, was used to obtain the masses in Col.~4, and the value 0.26 m$^2$ kg$^{-1}$, for  {\it diffuse ISM}, was used to obtain the masses in Col.~5. The $\kappa_{450}$ for {\it diffuse ISM}, in Col.~5, is the best match to the sub-mm dust mass obtained from \ICO\ data, in Col.~6.

\begin{table*}
  \caption{ Masses from \ICO\ Emission}
  \label{masses1}
  \centering
  \begin{tabular}{|c|c|c|c|c|c|}
    \hline
          (1)   &    (2)   &      (3)  &   (4)&     (5)& (6)        \\
  \ICO\         & Radius   & Spherical &H$_2$ &    Gas &Dust      \\
   Cloud        & (FWHP)   & Radius    &density & Mass & Mass \\
         Name   & (\arcsec)& (pc)      & (cm$^{-3}$)& M$_\odot$&M$_\odot$\\
    \hline
      A         &   40     & 0.43       &1000   &22&0.2       \\
      B         &60        &0.65        &1000   &77 &0.8      \\
      C         &20        &0.22        &400    &1  &0.01     \\
\hline
      E         &50        &0.54        &1000   &44 &0.4  \\
\hline

      F         &65        &0.70        &1000   &100  & 1.0    \\
      G         &85        &0.89        &2000   &400   & 4.0         \\
      H         &40        &0.42        &$<$800 &$<$17& 0.2 \\
    \hline
  \end{tabular}
\end{table*}

\begin{table*}
  \caption{Dust Masses from both 450~$\mu$m dust and \ICO\ emission}
  \label{masses}
  \centering
  \begin{tabular}{|c|c|c|c|c|c|}
    \hline
          (1)   &            (2)            &          (3)   &(4) & (5)  &(6)    \\
\ICO\ & Dust&Dust Mass &Dust Mass & Dust Mass  & Dust Mass from \ICO\   \\
 Cloud& Cloud &(amorphous) &(newly formed)&(diffuse ISM)&                   \\ 
Name  & Name  &($\kappa_{450}$=1.5 m$^2$kg$^{-1}$)& ($\kappa_{450}$=0.88 m$^2$kg$^{-1}$)& ($\kappa_{450}$=0.26 m$^2$kg$^{-1}$)  &      \\

 &       &M$_\odot$& M$_\odot$& M$_\odot$ & M$_\odot$           \\
    \hline
      A, B, C   & 1&           0.2&0.3& 1.2           &          1.0         \\
      E         & 3&           0.2&0.3& 1.0           &          0.4         \\

      F, G, H   & 5&           0.7 &1.2  &  4.0       &          5.1         \\
    \hline
  \end{tabular}
\end{table*}

\section{Discussion}

\subsection{Dust toward and in Cas~A}
We have argued that at least  one half of the mass of 
dust measured toward Cas~A is in foreground interstellar clouds. Thus the dust mass quoted by
Dunne \etal(2003) should be reduced by at least a factor of 2. Taking
the favored value from Dunne \etal (2003), we find
that Cas~A produced at most 1.5~M$_\odot$ of dust if we use the sub-mm dust absorption coefficient favored by Dunne \etal (2003). 
Some additional support for identifying the dust with CO clouds is
obtained from a comparison of gas and dust temperatures. As
reported by Dunne \etal(2003), most of the mass contained in
dust is cold, with a temperature of 18 K. This is
consistent with the peak temperatures of the CO measured toward
Cas~A, $\sim$20~K (Wilson \etal \cite{wil}) and the rotational
temperature of \AMM, about 18 K (Batrla \etal \cite{bat}. 

The \AMM\ molecule is easily dissociated, whereas the CO is much more robust, and so is present throughout molecular clouds. Thus the \AMM\ should be present only in cores of clouds; the VLA measurement of \AMM\ absorption toward Cas A (Gaume \etal 1994) directly shows this is so.  From the equality of the CO and \AMM\ temperatures, the clouds are isothermal. In addition, the dust and gas temperatures are very similar.  In most cases, this indicates a
close coupling of dust and gas, implying densities of order
10$^4$~\percc, higher than the average for \ICO\ clouds in
Table~1, 400 \percc\ to 2000 \percc. This may indicate denser cores in these clouds. The average minimum dust temperature found for extended cirrus clouds (Lagache \etal 1998) is 17.5 K. If the clouds are very clumpy, the heating may extend to the interior and thus allow us to explain the equality of gas temperatures in spite of the relatively small \MOLH\ densities. If the cores of these clouds are heated by photons, there is an unsolved problem of the dissociation of \AMM. In a comparison of large and small galactic clouds Turner (1995) has found the abundance of \AMM\ in clouds similar to those measured toward Cas A is too large compared to his model predictions, so additional production mechanisms for \AMM\ are needed.

\subsection{General Comments}

Morgan \& Edmunds (\cite{mo2}) have pointed out that there are a
large number of dusty galaxies measured with redshift $z=3$. These
authors have produced a model for dust production which shows significant differences when
compared to the model of Dwek (\cite{dw2}), showing that
predictions are rather parameter dependent. However, since
the production of dust in Red Giant stars is slow, this production path would
require a very large star formation rate. Thus, supernovae may be the production sites for dust in the early universe.  However, a number of
questions remain. The first point (raised by Dunne
\etal(\cite{dun})), is the following. Using a sub-mm dust absorption coefficient
which is characteristic of the diffuse ISM gives rise to a very
large dust mass. To arrive at their favored dust mass estimate, Dunne
\etal(\cite{dun}) used an absorption coefficient for amorphous
dust or for dust aggregates. Following this approach, but in much more extreme way,  
Dwek (\cite{dw1}) attributes the sub-mm dust thermal emission from SNe
found by Dunne \etal(\cite{dun}) and Morgan \etal(\cite{mo1}) in
terms of emission by conducting needles. According to Dwek
(\cite{dw1}), these are more efficient emitters of sub-mm thermal
continuum radiation. If so, the dust
mass would be reduced by factors of 10$^3$ or more. Another, less extreme, model of dust emission from the Kepler SNR is that of Contini (\cite{cont}). In this model only very large grains survive in the harsh SNR environment. These are more efficient emitters in the mm/sub-mm wavelength range, so the total mass of dust would be a factor of 10 smaller, but this model can fit all of the existing broadband dust measurements. From a comparison of 24 $\mu$m, 70 $\mu$m and sub-mm dust emission images, Hines \etal (2004) note that if the dust emitting at sub-mm wavelengths resides within the SNR, it must have very different properties from the dust emitting in the mid-IR. 

If the Dwek 
supposition is correct, the spectral index of the SNR dust must be
different from that of typical interstellar dust. Then, with $\sim$15\arcsec\ resolution far-IR
images, for example with the PACS imaging photometer instrument on
Herschel or the HAWC far IR bolometer camera on SOFIA, one should
be able to map the spectral indices of thermal dust emission across Cas~A and and toward the nearby clouds. If one could distinguish
between interstellar dust and dust inside the SNR, this would be evidence for different properties. Such a finding  would have very far-reaching consequences for the mass of dust
produced by SNRs. Another future line of investigation involves a study of the properties of the interstellar clouds near the edges of the Cas A SNR. The presence of line wings in CO or a change of excitation in \ICO\ might indicate an interaction of the clouds and the SNR. Although not directly related to Cas A, an image of a larger region near the SNR in both \ICO\ lines and sub-mm dust emission would allow a more complete determination of cloud properties.

\section{Conclusions}

From a comparison of the distribution of the image of integrated
J=1--0 \ICO\ line emission (Wilson \etal \cite{wil}) with the
image of 450 $\mu$m dust emission (Dunne \etal \cite{dun}), we find
positional agreement for 3 of 6 maxima. One sub-mm dust peak, \lq2\rq\ was not mapped in \ICO, while another, \lq6\rq\ lies outside the forward shock of the SNR. The third  prominent dust emission peak, \lq4\rq\ does
not show \ICO\ emission. There is \HI absorption at this position, but the column density of \HI may not be sufficient to account for the dust emission, so this may be inside the SNR. Both the \ICO\ and sub-mm dust images show that the emission is concentrated south of the center of Cas~A, while the synchrotron emission is a rather symmetric ring-shaped structure. There is good
agreement between the masses of the sub-mm dust maxima coincident with
the \ICO\ peaks and also good agreement between the dust
temperature and the peak molecular line temperatures from CO and
\AMM. We conclude that {\it at least} one-half of the dust emission
measured toward Cas~A arises from molecular clouds toward, but
{\it not} inside the SNR. From the conversion favored by 
Dunne \etal(\cite{dun}) and our corrections for dust in line-of-sight molecular interstellar clouds, we find that the  mass of dust inside the SNR is {\it at most} 1.5~M$_\odot$. 

\noindent
{\bf acknowledgement} We thank C.~M.~Walmsley and a referee for valuable comments.

\end{document}